\documentclass{pasj00}
\usepackage[dvips]{graphics}
\begin{document}
\SetRunningHead{C.\ Akizuki and J.\ Fukue}{ADAF with Toroidal Magnetic Fields}
\Received{yyyy/mm/dd}
\Accepted{yyyy/mm/dd}
%adaf2004.tex 2004 0911
%referee      2004 
%editing      2004 

\title{Self-Similar Solutions for ADAF with Toroidal Magnetic Fields}

%%% begin:list of authors
\author{Chizuru \textsc{Akizuki} and Jun \textsc{Fukue}} %   \thanks{}}
\affil{Astronomical Institute, Osaka Kyoiku University, 
Asahigaoka, Kashiwara, Osaka 582-8582}
\email{j059337@ex.osaka-kyoiku.ac.jp, fukue@cc.osaka-kyoiku.ac.jp}

%\author{B-Firstname \textsc{B-Familyname}}
%\affil{B-Address of Institute}\email{bbbbb@xxx.xxx.xx.xx}
%\and
%\author{C-Firstname {\sc C-Familyname}}
%\affil{C-Address of Institute}\email{ccccc@xxx.xxx.xx.xx}
%%% end:list of authors

%% `\KeyWords{}' always has to be placed before `\maketitle'.
 %Do NOT move this preamble from here!

\maketitle

%\newpage

\begin{abstract}
We examined the effect of toroidal magnetic fields on a viscous gaseous disk  
around a central object under an advection dominated stage.
We found self-similar solutions for radial infall velocity,
rotation velocity, sound speed,
with additional parameter $\beta$ [$=c_{\rm A}^2/(2c_{\rm s}^2)$],
where $c_{\rm A}$ is the Alfv\'en speed and
$c_{\rm s}$ is the isothermal sound speed.
Compared with the non-magnetic case, in general
the disk becomes thick due to the magnetic pressure,
and the radial infall velocity and rotation velocity become fast.
In a particular case, where the magnetic field is dominant,
on the other hand,
the disk becomes to be magnetically supported, and
the nature of the disk is significantly different from
that of the weakly magnetized case.
\end{abstract}

\section{Introduction}

Accretion-disk models have been extensively studied
during these three decades  (see Kato et al. 1998 for a review),
and several types of models were proposed.

One is the classical standard disk,
where the accretion rate is subcritical,
the disk is geometrically thin and optically thick,
Keplerian rotating with negligible infall velocity,
and truncated at the marginally stable orbit.
Such a standard disk was studied
for the Newtonian case (e.g., Shakura, Sunyaev 1973) and
for the relativistic case (e.g., Novikov, Thorne 1973).

When the accretion rate is so low,
the disk is supposed to be in the optically-thin 
advection-dominated state,
where 
the disk is optically thin with insufficient cooling
and extends down to the surface of the central object
 (e.g., Ichimaru 1977; Narayan, Yi 1994).
Such an optically-thin advection-dominated accretion flow (ADAF)
has been examined
for the Newtonian case
 (e.g., Narayan, Yi 1994, 1995),
under the pseudo-Newtonian potential
 (e.g., Narayan et al. 1997), and
for the Kerr metric
 (Gammie, Popham 1998; Popham, Gammie 1998).

When the accretion rate is so high, on the other hand,
the disk is supposed to be in the optically-thick
advection-dominated state,
where 
the disk is optically thick with photon trapping
and extends down to the surface of the central object
 (e.g., Abramowicz et al. 1988; Eggum et al. 1988).
Such an optically-thick ADAF (slim disk)
has been examined
for the Newtonian case
 (e.g., Abramowicz et al. 1988; Watarai, Fukue 1999; Fukue 2000),
under the pseudo-Newtonian potential
 (e.g., Szuszkiewicz et al. 1996; Artemova et al. 2001), and
for the Kerr metric
 (e.g., Abramowicz et al. 1996;
 Igumenshchev et al. 1998; Beloborodov 1998).

Disk models with outflows were also proposed.
For an example, 
Blandford and Begelman (1999) proposed
an adiabatic inflow-outflow solution (ADIOS)
for an optically-thin state,
where the wind is driven by the hot gas pressure
(see also Blandford, Begelman 2004).
While, for an optically thick state,
Fukue (2004) proposed a critical accretion disk,
where the wind is driven by the radiation pressure
(see also Lipunova 1999; Kitabatake et al. 2002).

Also investigated was a convection-dominated accretion flow (CDAF),
where the disk is convective in the radial direction
(Narayan et al. 2000; Quataert, Gruzinov 2000).
In the Newtonian case,
self-similar solutions were found
when convection moves angular momentum inwards.

The effect of magnetic fields 
has also been studied, in particular, in relation to 
magneto-rotational instability (MRI) (Balbus, Hawley, 1998),
and an accretion flow in which the magnetic forces dominate
over the thermal and radiation forces is called
a magnetically-dominated accretion flow (MDAF)
(e.g., Meier 2005; see also Shadmehri, Khajenabi 2005).
However, the intermediate case,
where the magnetic force is comparable to other forces,
has not been studied well.
Moreover, the mass loss and escape of magnetic fields
for MDAF were not considered.

When the sufficient amount of chaotic magnetic fields
are created by, e.g., MRI and
the ohmic heating operates effectively,
electrons are heated as well as ions,
and ADAF conditions would not be realized
(e.g., Bisnovatyi-Kogan, Lovelace 1997, 2001).
We do not consider, however, such chaotic fields
as well as the ohmic heating in the present analysis.
Incidentally in this paper, we consider 
the effect of global toroidal fields on the advection-dominated disk,
taking into accout the mass loss as well as the field escape.

In the next section 
the basic equations for the present purpose are presented.
In sections 3 and 4, self-similar solutions are shown.
We discuss several properties of the present model in section 5.
The final section is devoted to concluding remarks.

\section{Basic Equations}

Let us suppose a gaseous disk rotating around a 
Schwarzschild black hole of mass $M$.
The disk is assumed to be in an advection-dominated state,
where the viscous heating is balanced with the advection cooling. 
In a cylindrical coordinates ($r, \varphi, z$), 
we vertically integrate the flow equations.  
Furthermore, the flow is assumed to be steady and axisymmetric 
($\partial/\partial t = \partial/\partial\varphi=0$), and all flow 
variables are a function only of $r$.
We ignore the general relativistic effects and use Newtonian gravity.
We adopt an $\alpha$ prescription (e.g., Shakura, Sunyaev 1973). 
As magnetic fields, we consider only toroidal fields $B_\varphi$.

For such a disk, the continuity equation with mass loss is
\begin{equation}
   \frac{1}{r} \frac{d}{dr} (r\Sigma v_r) = 2\dot{\rho} H,
   \label{continuity}
\end{equation} 
where $v_r$ is the radial infall velocity, 
$\dot{\rho}$ the mass-loss rate per unit volume,
$H$ the disk half-thickness, 
and $\Sigma$ the surface density, which is defined as 
$\Sigma \equiv 2 \rho H$, $\rho$ being density.

The radial momentum equation is 
\begin{equation}
   v_r \frac{dv_r}{dr} =\frac{v_\varphi^2}{r} - \frac{GM}{r^2}
      -\frac{1}{\Sigma}\frac{d}{dr}
     \left( \Sigma c_{\rm s}^2 \right)-\frac{c_{\rm A}^2}{r} 
     - \frac{1}{2\Sigma}\frac{d}{dr} \left( \Sigma c_{\rm A}^2 \right),
\label{momentum}
\end{equation}
where $v_\varphi$ is the rotation velocity, 
$c_{\rm s}$ the isothermal sound speed,
which is defined as $c_{\rm s}^2 \equiv p_{\rm gas}/\rho$,
 $p_{\rm gas}$ being the gas pressure,
and $c_{\rm A}$ is Alfv\'{e}n speed,
which is defined as $c_{\rm A}^2 \equiv B_{\rm \varphi}^2/(4\pi\rho)
=2p_{\rm mag}/\rho$,
$p_{\rm mag}$ being the magnetic pressure.
On the right-hand side of equation (\ref{momentum}),
 the third term implies the pressure gradient force, while  
the magnetic force is designated as the fourth and fifth terms.

As an alpha prescription, we consider two extreme cases:
the $r\varphi$-component of the viscous stress tensor is 
proportional to the {\it gas pressure} or
to the {\it total pressure}.
In addition, instead of
$t_{r\varphi} = \eta r d\Omega/dr = - \alpha p$,
where $\eta$ is the viscosity and 
$\alpha$ is the viscous parameter (Shakura, Sunyaev 1973),
we adopt the form of
$\nu = \Omega_{\rm K}^{-1}\alpha (p/\rho)$,
where $\nu$ is the kinematic viscosity (Narayan, Yi 1994).
That is, two cases we consider are
\begin{equation}
   \eta = \rho \nu = \left\{
\begin{array}{ll}
   \Omega_{\rm K}^{-1} \alpha p_{\rm gas} & ~~~~~{\rm case~1}, \\
   \Omega_{\rm K}^{-1} \alpha (p_{\rm gas}+p_{\rm mag}) & ~~~~~{\rm case~2}.
\end{array}
   \right.
\label{alpha}
\end{equation}

Hence, the angular momentum transfer equation is
\begin{equation}
   r\Sigma v_r \frac{d}{dr} \left( r v_\varphi \right)
   = \frac{d}{dr} \left( \frac{\alpha \Sigma c_{\rm s}^2 r^3}
    {\Omega_{\rm K}} \frac{d\Omega}{dr} \right) ~~~~~{\rm case~1},
\label{angular}
\end{equation}
where $\Omega$ ($=v_\varphi/r$) is the angular speed,
and $\Omega_{\rm K}$ ($ =\sqrt{GM/r^3} $) 
the Keplerian angular speed.

In both cases,
the hydrostatic balance in the vertical dirction is integrated to
\begin{equation}
   \frac{GM}{r^3} H^2 = c_{\rm s}^2 \left[1 + \frac{1}{2} 
   \left(\frac{c_{\rm A}}{c_{\rm s}}\right)^2 \right]
   =(1+\beta)c_{\rm s}^2.
\label{hydrostatic}
\end{equation}
Here, we introduce the parameter $\beta$ by
\begin{equation}
   \beta \equiv \frac{p_{\rm mag}}{p_{\rm gas}}
   =\frac{1}{2}\left(\frac{c_{\rm A}}{c_{\rm s}}\right)^2,
\end{equation}
which is the degree of magnetic pressure to gas pressure.
We assume this ratio is constant through the disk.

The energy equation becomes
\begin{equation}
\frac{v_r}{\gamma -1} \frac{dc_{\rm s}^2}{dr} +
 \frac{c_{\rm s}^2}{r} \frac{d}{dr}\left(rv_{\rm r} \right) 
 = f \frac{{\alpha c_{\rm s}^2} r^2}{\Omega_{\rm K}} 
   \left(\frac{d\Omega}{dr} \right)^2 ~~~~~{\rm case~1},
\label{energy}
\end{equation}
where $\gamma$ is the ratio of specific heats.
The advection parameter $f$ measures the degree 
to which the flow is advection-dominated (Narayan, Yi 1994),
 and is assumed to be constant.

Finally, since we consider only the toroidal field,
the induction equation with field escape can be written as
\begin{equation}
   \frac{d}{dr}\left(v_r B_{\rm \varphi} \right) = \dot{B}_\varphi,
\label{induction}
\end{equation}
where $\dot{B}_\varphi$ is the field escaping/creating rate
due to magnetic instability or dynamo effect.
This induction equation is rewritten as 
\begin{equation}
   v_r\frac{dc_{\rm A}^2}{dr}+c_{\rm A}^2\frac{dv_r}{dr}
   -\frac{c_{\rm A}^2 v_r}{r} = 
   2c_{\rm A}^2\frac{\dot{B}_\varphi}{B_\varphi}
   -c_{\rm A}^2 \frac{2\dot{\rho}H}{\Sigma}.
\end{equation}

\section{Self-Similar Solutions for Case 1}

We first consider an advection-dominated accretion flow,
 including toroidal magnetic fields,
for the case 1,
where the viscosity is assumed to be proportional to the gas pressure.
Under the self-similar treatment by Narayan and Yi (1994),
 the velocities are assumed to be expressed as follows,
\begin{eqnarray}
    v_r(r) &=& -c_1 \alpha \sqrt{\frac{GM}{r}},
\label{vr}
\\
    v_\varphi(r) &=& c_2 \sqrt{\frac{GM}{r}},
\label{vp}
\\
    c_{\rm s}^{2}(r)  &=& \frac{p}{\rho} = c_3 \frac{GM}{r},
\label{cs}
\\
    c_{\rm A}^{2}(r)  &=& \frac{B_\varphi^2}{4\pi \rho}
    = 2\beta c_3 \frac{GM}{r},
\label{cA}
\end{eqnarray}
where coefficients $c_1$, $c_2$, and $c_3$ are determined later.

In addition, the surface density $\Sigma$ is assumed to be a form of 
\begin{eqnarray}
    \Sigma(r) &=& \Sigma_0 r^{\rm s},
\label{sigma}
\end{eqnarray}
where $\Sigma_0$ and $s$ are constants.
Then, in order for the self-similar treatment to be valid,
the mass-loss rate per unit volume and 
the field escaping rate must have the following form,
\begin{eqnarray}
    \dot{\rho} &=& \dot{\rho}_0r^{s-5/2},
\label{rho}
\\
    \dot{B}_\varphi &=& \dot{B}_0r^{({s-5})/{2}},
\label{be}
\end{eqnarray}
where 
$\dot{\rho}_0$ and $\dot{B}_0$ are constants.

Using these solutions, from the continuity, momentum, angular momentum, 
hydrostatic, energy, and induction equations
[(\ref{continuity}), (\ref{momentum}), (\ref{angular}), 
(\ref{energy}), and (\ref{induction})],
we can obtain the following relations,
\begin{eqnarray}
  \dot{\rho}_0 &=& - \left(s+\frac{1}{2} \right) 
  \frac{c_1 \alpha \Sigma_0}{2} \sqrt{\frac{GM}{(1+\beta)c_3}},
\label{massloss}
\\
  -\frac{1}{2} c_1^2 \alpha^2 &=& c_2^2 -1 -[s-1+\beta(s+1)] c_3, 
\label{relation1}
\\
  c_1 &=& 3(s+1)c_3,
\label{relation2}
\\
  H/r &=& \sqrt{(1+\beta) c_3},
\label{height}
\\
  c_2^2 &=& \frac{3-\gamma}{\gamma-1} \frac{2}{9f} c_1
       \equiv \epsilon''c_1,
\label{relation3}
\\
  \dot{B}_0 &=& \frac{2-s}{2} c_1 \alpha GM 
   \sqrt{4\pi\Sigma_0 \frac{\beta c_3}{\sqrt{(1+ \beta)c_3}}}.
\label{escape}
\end{eqnarray}

As is easily seen from equation (\ref{height}), in general
the disk thickness becomes large due to the magnetic pressure
for the weakly to moderately magnetized cases of $\beta \sim 1$.
In addition,
for $s=-1/2$, there is no mass loss,
while there exists mass loss (wind) for $s > -1/2$.
On the other hand,
the ecape and creation of magnetic fields are balanced each other for $s=2$.

Considering these properties,
we found two types of solutions under the present treatment.
First is the {\it conical disk without accretion}
in the limit of a small alpha,
where the conical disk is supported by rotation, the gas pressure,
and the magnetic pressure.
In this case, $\dot{\rho}_0=0$ and $\dot{B}_0=0$.

Another type is the {\it advection disk under dynamo action}
in the case of a finite alpha,
where the toroical magnetic field is injected (and escaping),
due to the action of the dynamo effect and instability.
In this case,
$\dot{\rho} \propto r^{s-5/2}$ and $\dot{B}_\varphi \propto r^{(s-5)/2}$,
in particular,
in the case of no wind ($s=-1/2$),
$\dot{\rho} =0$ and $\dot{B}_\varphi \propto r^{-11/4}$.

Now, the relations (\ref{relation1}), (\ref{relation2}),
and (\ref{relation3}) determine the constants $c_i$.

When $\alpha=0$ at the no-accretion limit,
 these coefficients become  
\begin{eqnarray}
  c_1 &=& \frac{3}
   {\frac{\displaystyle 1-s}{\displaystyle 1+s}-\beta+3\epsilon''},
\\
  c_2^2 &=& \frac{3\epsilon''}
   {\frac{\displaystyle 1-s}{\displaystyle 1+s}-\beta+3\epsilon''},
\\
  c_3 &=& \frac{1}
   {(1+s)\left(\frac{\displaystyle 1-s}{\displaystyle 1+s}-\beta+3\epsilon''\right)},
\end{eqnarray}
where
\begin{equation}
   \epsilon'' = \frac{2}{9}\frac{3-\gamma}{\gamma-1}\frac{1}{f}.
\end{equation}

On the other hand, when $\alpha \neq0$, 
these coefficients become
\begin{eqnarray}
  c_1 &=& \frac{1}{3\alpha^2}
  \left[
  \sqrt{\left(\frac{1-s}{1+s}-\beta+3\epsilon''\right)^2+18\alpha^2}
  \right.
\nonumber \\
  && ~~~~~~\left.
   -\left(\frac{1-s}{1+s}-\beta+3\epsilon''\right)
  \right],
\\
  c_2^2 &=& \frac{\epsilon''}{3\alpha^2}
  \left[
  \sqrt{\left(\frac{1-s}{1+s}-\beta+3\epsilon''\right)^2+18\alpha^2}
  \right.
\nonumber \\
  && ~~~~~~\left.
   -\left(\frac{1-s}{1+s}-\beta+3\epsilon''\right)
  \right],
\\
  c_3 &=& \frac{1}{9(1+s)\alpha^2}
  \left[
  \sqrt{\left(\frac{1-s}{1+s}-\beta+3\epsilon''\right)^2+18\alpha^2}
  \right.
\nonumber \\
  && ~~~~~~\left.
   -\left(\frac{1-s}{1+s}-\beta+3\epsilon''\right)
  \right].
\end{eqnarray}

The parameters of the model are the ratio of the specific heats $\gamma$,
the standard viscous parameter $\alpha$, 
the magnetic pressure fraction $\beta$, 
the advection parameter $f$,
and the mass-loss parameter $s$.  
In the case without mass loss,
$ s = -1/2 $, as already stated.
In addition,
the effect of mass loss is similar to the advection effect (Fukue 2004).

Examples of coefficient $c_i$'s are shown in figure 1 
as a function of the advection factor $f$ for several 
values of $\beta$.
Other parameters are
$\gamma=4/3$, $\alpha=1$, and $s=-1/2$ (no wind).

\begin{figure}
  \begin{center}{
   \includegraphics[width=100mm]{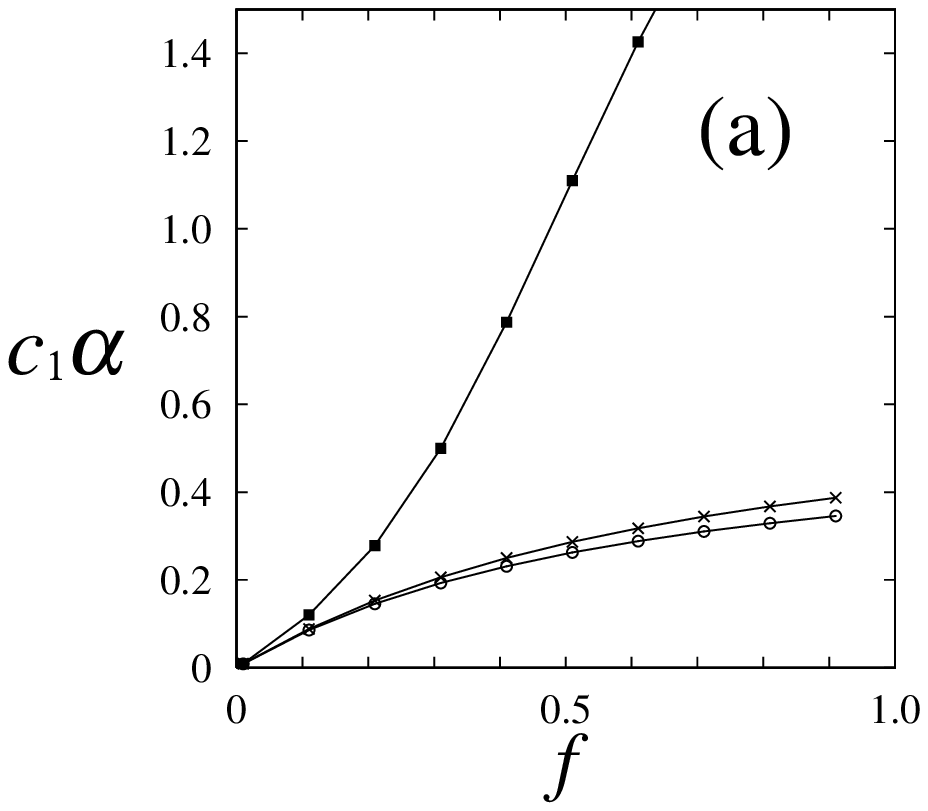}   
   \includegraphics[width=100mm]{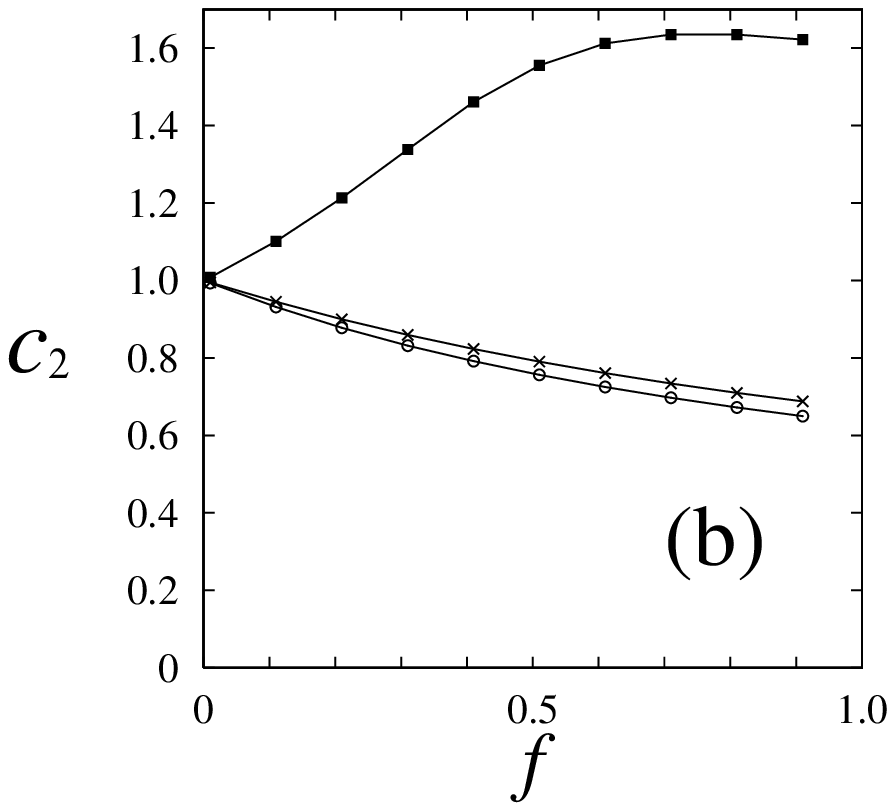}
   \includegraphics[width=100mm]{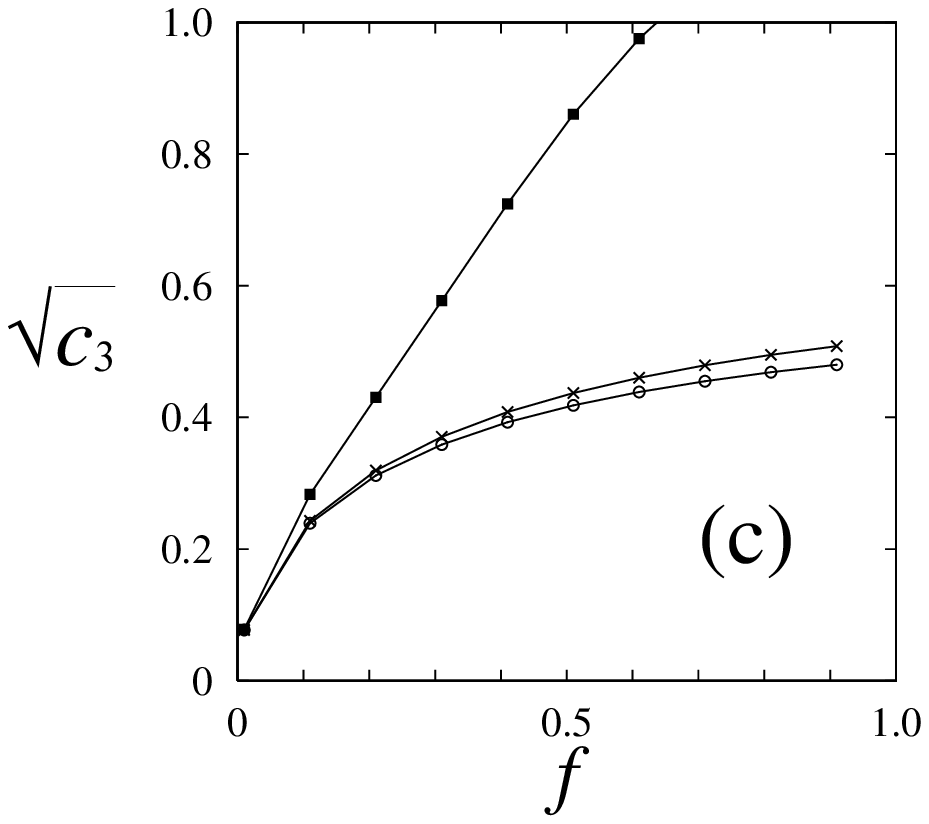}
}
\end{center} 
\caption{
Numerical coefficient $c_i$'s
as a function of the advection parameter $f$
for several values of $\beta$ in case 1:
(a) $c_1 \alpha$,
(b) $c_2$, and
(c) $\sqrt{c_3}$.
The disk density profile is set to be $s=-1/2$ (no wind),
the ratio of specific heats is set to be $\gamma=4/3$,
and the viscous parameter is $\alpha=1$.
The value of $\beta$ are 0, 1, and 10 from bottom to top
 of each curve in each panel.
}
\end{figure}

Figure 1a shows a coefficient $c_1 \alpha$,
which is a ratio of the radial infall velocity to the Keplerian one. 
Although the radial infall velocity is generally slower 
than the Keplerian speed ($ = \sqrt{{GM}/{r}} $),
it becomes large with $\alpha$ or $f$. 
In addition, the radial infall velocity becomes fast 
as the toroidal magnetic field becomes large. 
This is because,
in the case of $s=-1/2$,
the magnetic tension term dominates the magnetic pressure term
in the radial momentum equation (\ref{momentum}),
that assists the radial infall motion.

Figure 1b shows a coefficient $c_2$,
which is a ratio of the rotation velocity to the Keplerian one. 
As is seen in figure 1b,
the rotation velocity decreases as the advection is superior
in the weakly to moderately magnetized cases. 
This is because
the disk gas must rotate faster than the case without the magnetic field
due to the action of the magnetic tension. 
On the other hand, in the strongly magnetized case ($\beta \sim 10$),
the rotation speed becomes faster than the Keplerian speed
(super-Keplerian),
in order for the centrifugal force to be balanced with
the magnetic force.

Figure 1c shows a coefficient $\sqrt{c_3}$,
which is a ratio of the sound speed to the Keplerian one. 
As the advection parameter becomes large,
the sound speed as well as the disk thickness become large.
As for the effect of the toroidal magnetic field,
the sound speed also increases as the magnetic field becomes large.
In addition, as already stated,
the disk thickness becomes large due to the effect of 
the magnetic pressure.

%A difference of viscosity and an effect of 
%magnetic pressure appear conspicuously so 
%that advection all figure become superior. 

\section{Self-Similar Solutions for Case 2}

We next examine an advection-dominated accretion flow,
 including toroidal magnetic fields,
for the case 2,
where the viscosity is assumed to be proportional to the total pressure.
As is easily seen from equation (\ref{alpha}),
we obtain the self-similar solutions for case 2,
if we replace $\alpha$ in equations (\ref{angular}) and (\ref{energy})
by $\alpha (1+\beta)$.

Hence, the solutions at the no-accretion limit of $\alpha=0$ are 
\begin{eqnarray}
  c_1 &=& \frac{3(1+\beta)}
   {\frac{\displaystyle 1-s}{\displaystyle 1+s}-\beta+3\epsilon''},
\\
  c_2^2 &=& \frac{3\epsilon''}
   {\frac{\displaystyle 1-s}{\displaystyle 1+s}-\beta+3\epsilon''},
\\
  c_3 &=& \frac{1}
   {(1+s)\left(\frac{\displaystyle 1-s}{\displaystyle 1+s}-\beta+3\epsilon''\right)}.
\end{eqnarray}

On the other hand, the solutions for $\alpha \neq0$ become
\begin{eqnarray}
  c_1 &=& \frac{1}{3\alpha^2(1+\beta)} \times
\nonumber \\
  &&
  \left[
  \sqrt{\left(\frac{1-s}{1+s}-\beta+3\epsilon''\right)^2+18\alpha^2(1+\beta)^2}
  \right.
\nonumber \\
  && ~~~~~~\left.
   -\left(\frac{1-s}{1+s}-\beta+3\epsilon''\right)
  \right],
\\
  c_2^2 &=& \frac{\epsilon''}{3\alpha^2(1+\beta)^2} \times
\nonumber \\
  &&
  \left[
  \sqrt{\left(\frac{1-s}{1+s}-\beta+3\epsilon''\right)^2+18\alpha^2(1+\beta)^2}
  \right.
\nonumber \\
  && ~~~~~~\left.
   -\left(\frac{1-s}{1+s}-\beta+3\epsilon''\right)
  \right],
\\
  c_3 &=& \frac{1}{9(1+s)\alpha^2(1+\beta)^2} \times
\nonumber \\
  &&
  \left[
  \sqrt{\left(\frac{1-s}{1+s}-\beta+3\epsilon''\right)^2+18\alpha^2(1+\beta)^2}
  \right.
\nonumber \\
  && ~~~~~~\left.
   -\left(\frac{1-s}{1+s}-\beta+3\epsilon''\right)
  \right].
\end{eqnarray}

Examples of coefficient $c_i$'s are shown in figure 2 
as a function of the advection factor $f$ for several 
values of $\beta$.
Other parameters are
$\gamma=4/3$, $\alpha=1$, and $s=-1/2$ (no wind).

\begin{figure}
  \begin{center}{
   \includegraphics[width=100mm]{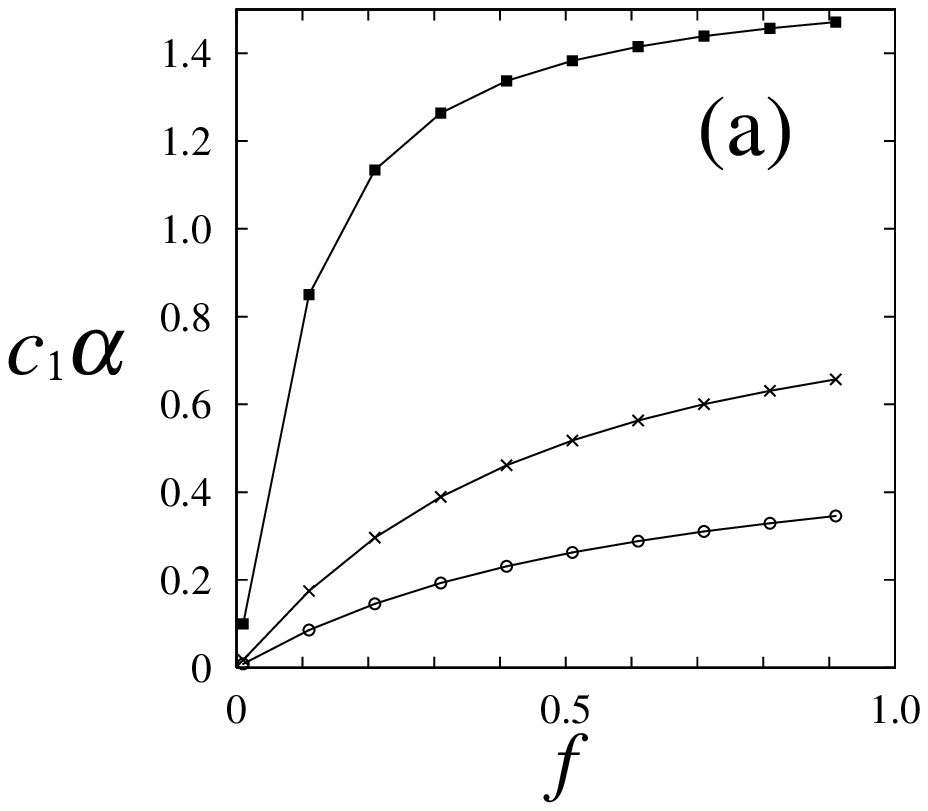}   
   \includegraphics[width=100mm]{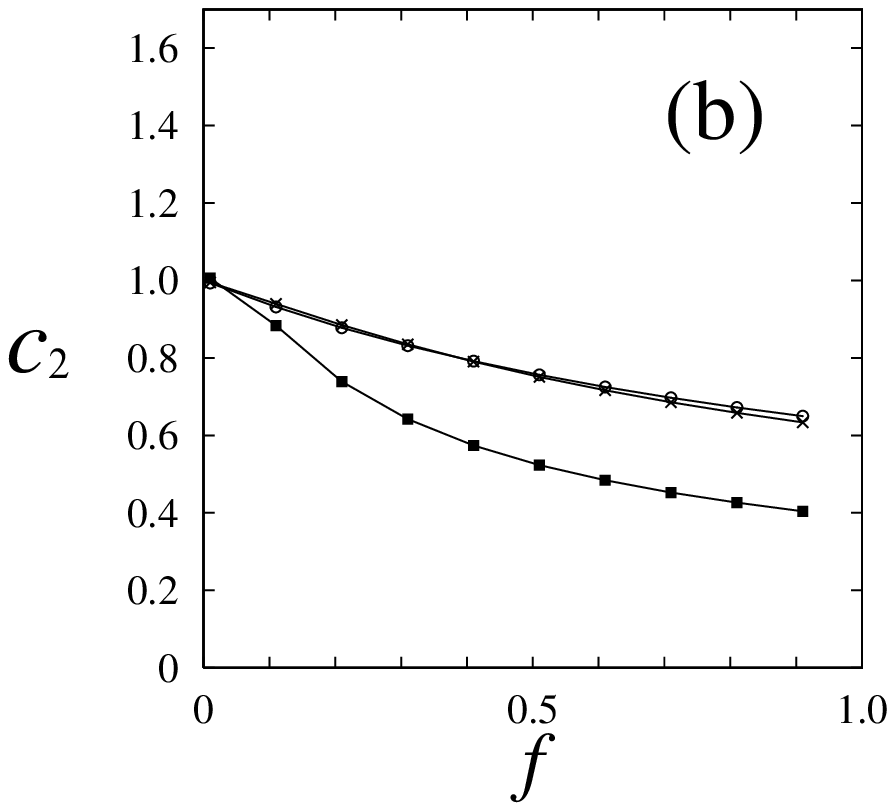}
   \includegraphics[width=100mm]{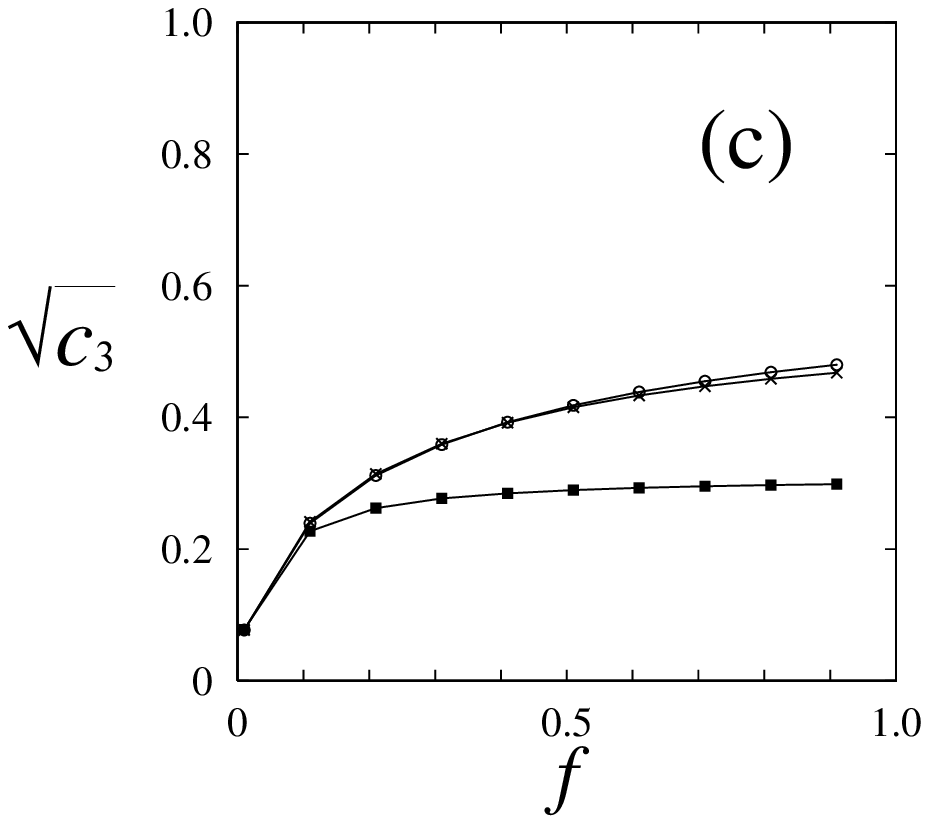}
}
 \end{center} 
\caption{
Numerical coefficient $c_i$'s
as a function of the advection parameter $f$
for several values of $\beta$ in case 2:
(a) $c_1 \alpha$,
(b) $c_2$, and
(c) $\sqrt{c_3}$.
The disk density profile is set to be $s=-1/2$ (no wind),
the ratio of specific heats is set to be $\gamma=4/3$,
and the viscous parameter is $\alpha=1$.
In (a), the value of $\beta$ are 0, 1, and 10 from bottom to top 
of each curve in the panel.
In (b) and (c), the value of $\beta$ are 0, 1, and 10 from top 
to bottom of each curve in each panel.
}
\end{figure}

In the weakly to moderately magnetized cases,
the behavior of the solutions for case 2
is qualitatively similar to those for case 1.
In the strongly magnetized case, however,
the behavior is drastically different between two cases,
as will be discussed in the next section.
It should be stressed here that both $c_2$ and $\sqrt{c_3}$ 
become small as $\beta$ increased.

It should be noted that in general case of viscosity:
\begin{equation}
   \eta=\Omega_{\rm K}^{-1} \alpha p_{\rm gas}^\mu
                       (p_{\rm gas}+p_{\rm mag})^{1-\mu},
\end{equation}
where $\mu$ is constant,
we easily obtain the solutions,
if we replace $\alpha$ in equations (\ref{angular}) and (\ref{energy})
by $\alpha (1+\beta)^{1-\mu}$.

\section{Discussion}

We briefly note several other properties
of the present self-similar solutions for ADAF
with toroidal magnetic fields.

At first, we show the temperature of the present
self-similar advection-dominated disk with toroidal magnetic fields.

In the optically thin case,
where the gas pressure is dominant,
the isothermal sound speed $c_{\rm s}$ is expressed as
\begin{equation}
   \frac{{\cal R}}{\bar{\mu}}T = c_{\rm s}^2 = \frac{GM}{r},
\end{equation}
where $T$ is the gas temperature,
${\cal R}$ the gas constant, and
$\bar{\mu}$ the mean molecular weight.
In this case, the temperature is expressed as
\begin{equation}
   T = c_3 \frac{c^2/2}{{\cal R}/\bar{\mu}} \frac{r_{\rm g}}{r}
     = 2.706 \times 10^{12} \frac{c_3}{r/r_{\rm g}} {\rm ~K},
\end{equation}
where $r_{\rm g}$ ($=2GM/c^2$)
is the Schwarzschild radius of the central object and $\bar{\mu}=0.5$.
This has a similar form with that of the non-magnetic case,
but the coefficient $c_3$ implicitly depends on the magnetic field.
That is,
the gas temperature increases as the magnetic field increases.

In the optically thick case,
where the radiation pressure dominant, on the other hand,
the sound speed is related to the radiation pressure
at the disk equator by
\begin{equation}
   \frac{(aT_{\rm c}^2/3)2H}{\Sigma} = c_{\rm s}^2 = \frac{GM}{r},
\end{equation}
where $a$ is the radiation constant,
and $T_{\rm c}$ the central temperature.
Then, in terms of the optical depth $\tau$ ($=\kappa \rho H$),
the effective temperature $T_{\rm eff}$ at the disk surface
is expressed as
\begin{equation}
   \sigma T_{\rm eff}^2 = \frac{1}{\tau} \sigma T_{\rm c}^4
   = \frac{3}{4} \sqrt{\frac{c_3}{1+\beta}} \frac{L_{\rm E}}{4\pi r^2},
\end{equation}
where
$L_{\rm E}$ ($=4\pi cGM/\kappa$)
is the Eddington luminosity of the central object.
Hence, the effective temperature becomes
\begin{eqnarray}
   T_{\rm eff} &=& 2.181 \times 10^7
   \left( \frac{c_3}{1+\beta} \right)^{1/8}
\nonumber \\
   &&
   \left( \frac{M}{10M_\odot} \right)^{-1/4}
   \left( \frac{r}{r_{\rm g}} \right)^{-1/2}
   {\rm ~K}.
\end{eqnarray}
In this case,
the effective temperature very weakly depends
on the strength of the magnetic field,
and therefore, it may be slightly changed
by the existence of toroidal magnetic fields.

The strength of toroidal magnetic fields becomes
\begin{equation}
  B_\varphi = 
   \sqrt{4\pi\Sigma_0 GM \frac{\beta c_3}{\sqrt{(1+ \beta)c_3}}}
   r^{s/2-1}.
\end{equation}
In the case without mass loss ($s=-1/2$),
the magnetic field depends on radius as
$B_\varphi \propto r^{-5/4}$.
This dependence is consistent with the usual MDAF
(Meier 2005; Shadmehri, Khajenabi 2005).

We briefly discuss the extremely magnetized case of large $\beta$ limit.
In case 1, where the viscosity is proportional to the gas pressure,
at the limit of large $\beta$,
the coefficients are approximately expressed as
\begin{eqnarray}
   c_1 &\sim& \frac{2}{3\alpha^2} \beta,
\\
   c_2^2 &\sim& \frac{2\epsilon''}{3\alpha^2} \beta,
\\
   c_3 &\sim& \frac{2}{9(1+s)\alpha^2} \beta.
\end{eqnarray}
In this case, $c_1$ could highly exceed unity,
which means that the gas dynamically infalls with super-Keplerian speed
due to the magnetic force.
Similary, the gas rotates with super-Keplerian speed in this case.

In case 2, where the viscosity is proportional to the total pressure,
at the limit of large $\beta$,
the coefficients are approximately expressed as
\begin{eqnarray}
   c_1 &\sim& \frac{\sqrt{1+18\alpha^2}+1}{3\alpha^2},
\\
   c_2^2 &\sim& \frac{\epsilon'' c_1}{\beta},
\\
   c_3 &\sim& \frac{c_1}{3(1+s)\beta}.
\end{eqnarray}
Hence, in this case, $c_1$ becomes ceiling,
while $c_2$ and $c_3$ become sufficiently small.
These properties are qualitatively consistent with
the results by Oda et al. (2005).

Finally, in general case, 
where $\nu \propto p_{\rm gas}^\mu (p_{\rm gas}+p_{\rm mag})^{1-\mu}$,
at the limit of large $\beta$ and for $0 < \mu < 1$,
the coefficients are approximately expressed as
\begin{eqnarray}
   c_1 &\sim& \frac{2}{3\alpha^2} \beta^\mu,
\\
   c_2^2 &\sim& \frac{\epsilon''}{3\alpha^2} \beta^{-1+2\mu},
\\
   c_3 &\sim& \frac{1}{9(1+s)\alpha^2} \beta^{-1+2\mu}.
\end{eqnarray}
This means that in the magnetically dominated disk of large $\beta$
the nature of the disk is drastically changed at $\mu=1/2$.
For $\mu>1/2$, the gas rotates with super-Keplerian speed, 
and the disk thickness becomes large,
while the quantities are very small for $\mu<1/2$.

Here, we briefly mention on
the magneto-rotational instability (MRI)
in relation to the above magnetically supported disk.
The structure with a strong pure toroidal field would be unstable 
because of a variant of MRI,
especially, in the interior of stars and accretion disks
(Tayler 1973; Spruit 2002; Akiyama et al. 2003;
Ardeljan et al. 2005; Balbus, Hawley 1992).
In the case of accretion disks, the growth time of MRI is 
about $1/\Omega$, where $\Omega$ is the angular speed of the disk,
while the inflow time is about $1/(\alpha \Omega)$
in the advection dominated accretion flow
considered in the present paper.
Hence, the toroidal magnetic field would infall as it grows via MRI,
and the strongly magnetized disk could be marginally stable.
In addition, from weakly to moderately magnetized cases, 
self-similar solutions with toroidal fields 
may be possible because the fields linearly grow.

Furthermore,
in numerical simulations of core-collapse supernovae
with magnetic fields
(Akiyama et al. 2003; Ardeljan et al. 2005),
the linear growth of toroidal fields is terminated
by the development of MRI,
leading to drastic acceleration in the growth of magnetic energy.
In the case of magnetized accretion disks, however,
a recent resistive MHD simulation of black hole accretion flows 
by Machida et al. (2005) shows that
MRI may be suppressed in the strongly magnetized case
due to the magnetic tension and pressure,
and the magnetically supported disk may be realized
(see also Oda et al. 2005).
Hence, 
we consider whether the disk with strong toroidal fields is 
unstable or not is still controversial.

\section{Concluding Remarks}

In this paper
we have examined the effect of toroidal magnetic fields
on  advection-dominated accretion flows.
We found self-similar solutions,
using the similarity  
technique in analogy to the self-similar solutions by Narayan and Yi (1994). 
Self-similar solutions would be valid in the region
 from about 10 Schwarzschild radii to about 1000 Schwarzschild radii,
in the case that the total disk size is about $10^5$ Schwarzschild radii. 
In the present intermediate case on the strength of the magnetic field,
the disk structure, e.g., the coefficient $c_i$'s,
slightly depends on the strength of the magnetic field.
However, the tendency of the effect of the todoidal field
would become clear in the present study;
it raises the infall velocity, the rotation velocity,
and the sound speed in the case that
the magnetic tension is dominant than the magnetic pressure,
and vice versa.

The properties of the present model, e.g., the magnetic field,
are similar to the results of other MDAF
(e.g., Meier 2005; Shadmehri, Khajenabi 2005; Oda et al. 2005),
although in the present model
the toroidal magnetic field is dominant.

This model was thought about a toroidal magnetic field in defiance of a relativity effect. 
If the central object is relativistic, the gravitational field should be changed. 
This problem is remained as a future work.

\vspace*{1pc}

The authors would like to thank
Dr. K. Watarai, Dr. Y. Kato, and Mr. H. Oda
for their valuable comments and discussions.
This work has been supported in part
by a Grant-in-Aid for the Scientific Research (15540235 J.F.) 
of the Ministry of Education, Culture, Sports, Science and Technology.

%The authors would like to thank Professor Y. Osaki and Dr. N. Shibazaki
%for their valuable comments and discussions.
%The author would like to thank an anonymous referee for valuable comments and u%seful suggestions.

%a Grant-in-Aid for Scientific Research (15540235 JF) 
%of the Ministry of Education, Culture, Sports, Science and Technology.

%\noindent

\end {document}